\documentclass[aps,pra,
twocolumn,
nofootinbib,
citeautoscript,
superscriptaddress,showkeys,showpacs]{revtex4-1} 
\usepackage{amsmath,amsfonts,amssymb}
\usepackage{color}
\usepackage[dvipsnames]{xcolor}
\definecolor{magnta}{HTML}{D000D0}
\usepackage{tabularx}
\usepackage{graphicx}
\usepackage[colorlinks=true, allcolors = blue]{hyperref}

\renewcommand{\vec}{\boldsymbol}
\newcommand{\mycite}[1]{[\onlinecite{#1}]}

\begin{document}

\title{Strain tuning and anisotropic spin correlations in iron-based
systems}

\author{Roland Willa}

\affiliation{Institute for Theory of Condensed Matter, Karlsruhe Institute of
Technology, Karlsruhe, Germany}

\author{Max Fritz}

\affiliation{Institute for Theory of Condensed Matter, Karlsruhe Institute of
Technology, Karlsruhe, Germany}

\author{J\"org Schmalian}

\affiliation{Institute for Theory of Condensed Matter, Karlsruhe Institute of
Technology, Karlsruhe, Germany}

\affiliation{Institute for Solid State Physics, Karlsruhe Institute of Technology,
Karlsruhe, Germany}

\begin{abstract}
Nematic order in the iron-based superconductors is closely tied to
a lattice distortion and a structural transition from tetragonal to
orthorhombic symmetry. External stress of the appropriate symmetry acts as a conjugate field of the nematic order parameter and can be utilized to detwin nematic domains but also smears an otherwise sharp nematic transition. On the other hand, applying stress in proper symmetry channels allows one to tune the
nematic phase transition. Recent experiments analyzed the stress-induced
changes of the nematic and magnetic phase transition temperature. Here we show
that the observed trends can be understood in terms of spin-induced
nematicity. The strain sensitivity is shown to be a fluctuation effect. The strong sensitivity to antisymmetric strain is a consequence of the anisotropic nature of the magnetic excitation spectrum. The formalism presented here can be naturally generalized to determine the strain-sensitivity of vestigial phases related to other magnetic states that have been observed in the iron-based systems, such as e.g.\ the spin-charge density wave and the spin-vortex crystals. 
\end{abstract}
\maketitle

\section{Introduction}
\begin{figure}[b]
\centering
\includegraphics[width = 0.45\textwidth]{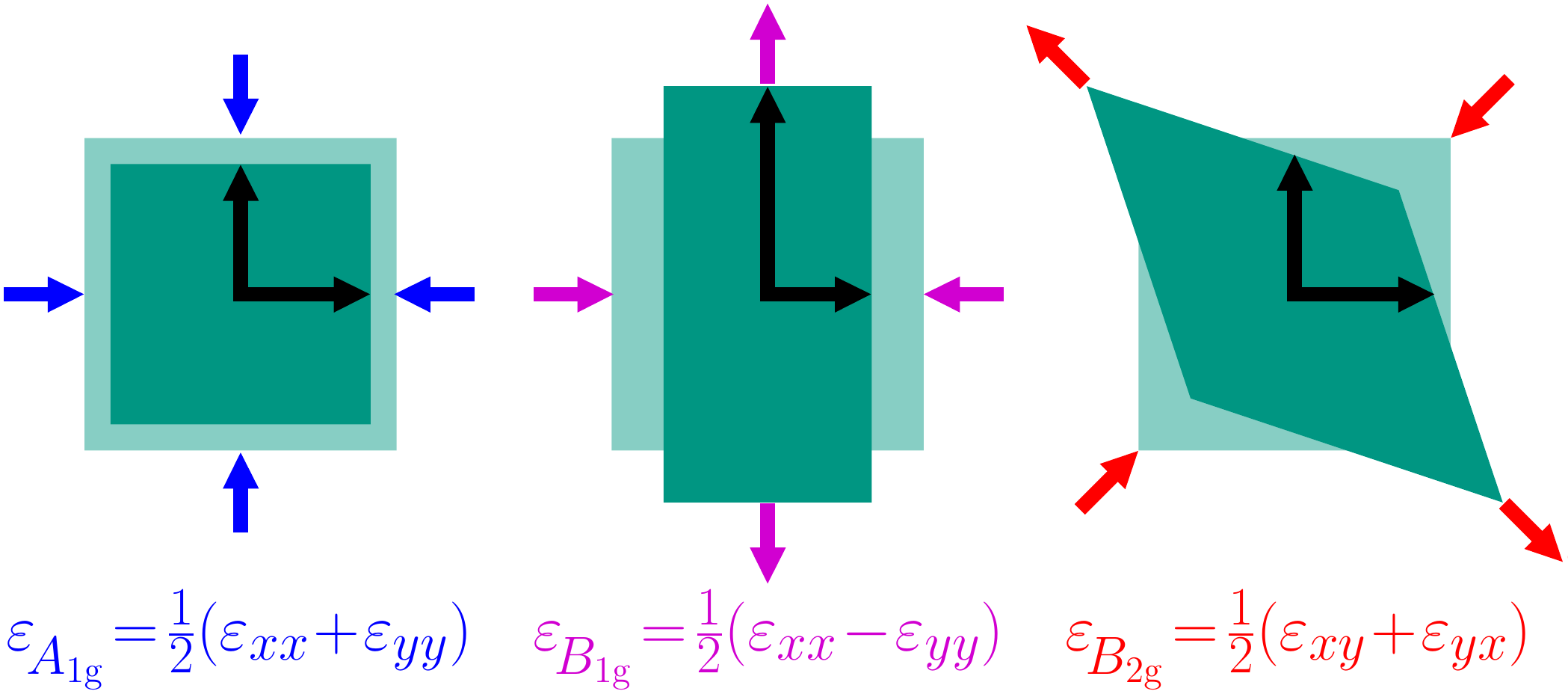}
\newcolumntype{L}[1]{>{\hsize=#1\hsize\raggedright\arraybackslash}X}%
\newcolumntype{R}[1]{>{\hsize=#1\hsize\raggedleft\arraybackslash}X}%
\newcolumntype{C}[1]{>{\hsize=#1\hsize\centering\arraybackslash}X}%
\begin{tabularx}{.9\columnwidth}{|L{1.72}|C{0.76}|C{0.76}|C{0.76}|}
\multicolumn{4}{c}{}\\[-.5em]
\hline
crystallographic axes & \multicolumn{3}{>{\hsize=\dimexpr2.1\hsize+4\tabcolsep+2\arrayrulewidth\relax}X|}{\qquad Symmetry sector}
\\\hline\hline
$\vec{e}_{x} \!=\! (100), \vec{e}_{y} \!=\! (010)$
& $\vec{\varepsilon_{A_{\mathrm{1g}}}}$
& $\vec{\varepsilon_{B_{\mathrm{2g}}}}$
& $\vec{\varepsilon_{B_{\mathrm{1g}}}}$
\\\hline
$\vec{e}_{x} \!=\! (110), \vec{e}_{y} \!=\! (\text{-}110)$
&    \textcolor{blue}{$\vec{\varepsilon_{A_{\mathrm{1g}}}}$}
& \textcolor{Magenta}{$\vec{\varepsilon_{B_{\mathrm{1g}}}}$}
&     \textcolor{red}{$\vec{\varepsilon_{B_{\mathrm{2g}}}}$}\\\hline
\end{tabularx}
\caption{Schematic action of different strain types on a square plaquette: Whereas the channels $A_{\mathrm{1g}}$ and $B_{\mathrm{2g}}$ preserve the nematic axes (black), the $B_{\mathrm{1g}}$ strain acts as a conjugate field, lifting the nematic degeneracy.
The nomenclature of the symmetry sectors for different basis vectors is clarified in the table.
}
\label{fig:sketch}
\end{figure}

Nematicity is a well-established state of electronic order in Fe-based superconductors \mycite{Fang2008, Xu2008, Si2008, Fernandes2010, Cano2010, Chuang2010, Chu2010, Lv2011, Fernandes2012a, Fernandes2012b, Chu2012, Liang2013, Stanev2013, Fernandes2014, Boehmer2014}. Nematic fluctuations were identified \mycite{Fernandes2010}
via a significant softening of the elastic shear modulus $C_{66}$,
a behavior that was interpreted in terms of spin-induced nematicity \mycite{Fang2008, Xu2008, Si2008, Fernandes2010, Fernandes2012a, Fernandes2012b, Fernandes2014}.
The latter can be understood as partially melted striped spin density-waves
that break rotational order, without breaking time-reversal symmetry.
Thus, within the spin-induced scenario, nematic order is caused by
fluctuations of $C_{2}$-symmetric single-$\vec{Q}$ spin order. The
strong coupling between elastic and nematic degrees of freedom has
been exploited in elastoresistivity measurements \mycite{Chu2012}
and three-point bending measurements \mycite{Boehmer2014}. It is rooted
in the fact that the Ising-nematic order parameter $\phi$ couples
bi-linearly to the corresponding strain $\varepsilon_{B_{\mathrm{1g}}}$.
Notice, by 'Ising'-order we mean a single-component scalar order-parameter that falls into the universality class of Ising systems.

\begin{figure}[tb]
\centering
\includegraphics[width = 0.35 \textwidth]{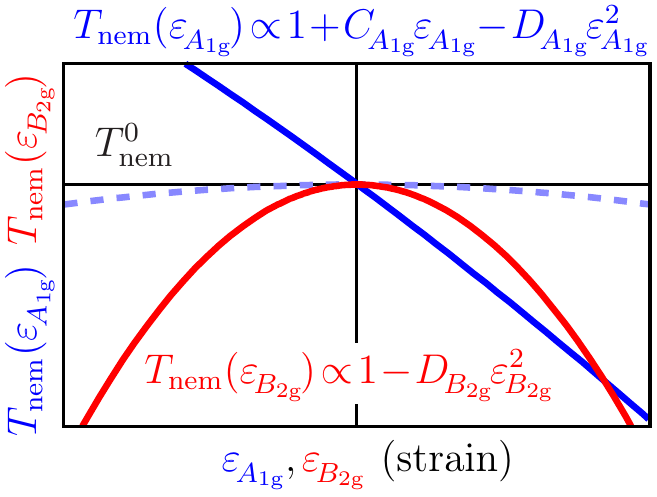}
\caption{Evolution of the nematic transition temperature upon applying strain in the $A_{\mathrm{1g}}$ (solid, blue) and $B_{\mathrm{2g}}$ (red) symmetry channels. The quadratic dependence in $A_{\mathrm{1g}}$ is highlighted when omitting the linear contribution (dashed, blue) and emphasizes the generic observation
that the quadratic correction in the $B_{\mathrm{2g}}$ sector is large, i.e. 
$D_{B_{\mathrm{2g}}} / D_{A_{\mathrm{1g}}} \gg 1$.
}
\label{fig:Tnem-evolution}
\end{figure}

Whereas this work shall focus on the nematicity associated with fluctuations of a spin-density wave order, the nematic order may also arise due to orbital or charge order \cite{Yamase2013, Onari2012}. One may further speculate that nematicity is at work in the hidden order observed in iridates \cite{Zhao2015}, or even at the origin of the pseudogap phase \cite{Fechner2016, Fradkin2010, Orth2019}. A transition of a parent disordered phase into a phase of electronic or spin nematicity that breaks a crystal symmetry is associated with a softening of elastic moduli and results in anisotropic transport responses. On the other hand a standard phonon-driven structural transition has almost identical anisotropic responses and may also facilitate the subsequent appearance of a magnetic or charge order. Resolving the problem of which mechanism is driving the transition has conclusively been answered in the pnictide materials, primarily thanks to elastoresistivity measurements \cite{Chu2010, Chu2012}.

Recently, Ikeda \emph{et al.}\ \mycite{Ikeda2018} investigated the
impact of strain on the nematic and magnetic phase transitions of $\mathrm{Co}$-doped $\mathrm{Ba}\mathrm{Fe}_{2}\mathrm{As}_{2}$ in different symmetry channels such as%
\footnote{In this paper we use a unit cell with iron-iron bonds along the coordinate axes. Thus, the irreducible representations $B_{\mathrm{1g}}$ and $B_{\mathrm{2g}}$ are interchanged if compared to Ref.\ \mycite{Ikeda2018}, see also Fig.\ \ref{fig:sketch}.}
$B_{\mathrm{2g}}$ and $A_{\mathrm{1g}}$, see Fig.\ \ref{fig:sketch}.
In distinction to $\varepsilon_{B_{\mathrm{1g}}}$ that couples directly to $\phi$,
strain in other symmetry channels will not wash out the nematic transition,
but shift its value. Thus, the nematic transition temperature $T_{\mathrm{nem}}$
remains sharply-defined. Among the key observations of Ref.\ \mycite{Ikeda2018}
are a quadratic \emph{suppression} of $T_{\mathrm{nem}}$ with antisymmetric
strain, i.e.
\begin{align}\label{eq:B2g-expectation}
   T_{\mathrm{nem}}=T_{\mathrm{nem}}^{0} (1 - D_{B_{\mathrm{2g}}} \varepsilon_{B_{\mathrm{2g}}}^{2}),
\end{align}
with $D_{B_{\mathrm{2g}}} > 0$, and a dominant linear variation for symmetric strain
\begin{align}\label{eq:A1g-expectation}
T_{\mathrm{nem}}=T_{\mathrm{nem}}^{0} (1 + C_{A_{\mathrm{1g}}} \varepsilon_{A_{\mathrm{1g}}} - D_{A_{\mathrm{1g}}} \varepsilon_{A_{\mathrm{1g}}}^{2}).
\end{align}
With regards to the quadratic response of the type $\propto \varepsilon^{2}$ it was further found that $D_{A_{\mathrm{1g}}} \!\ll\! D_{B_{\mathrm{2g}}}$, i.e. the change in the nematic transition temperature due to antisymmetric strain is stronger than due to symmetric strain.

In this paper we analyze the tuning of nematic order through to critical symmetric
and antisymmetric strain within the theory of spin-driven nematicity.
We demonstrate that the effect of strain is a fluctuation effect,
a behavior that is caused by the frustrated nature of the striped
magnetic order. We then show that the behavior observed in Ref.\ \mycite{Ikeda2018}
follows naturally within the approach of spin-driven nematicity. The
suppression of $T_{\mathrm{nem}}$ by strain is shown to be a consequence of strong
classical magnetic fluctuations. As such, strain-enhanced magnetic fluctuations provide a second route to suppress $T_{\mathrm{nem}}$, in addition to increasing quantum fluctuations, see Ref.\ \cite{Maharaj2017}. The finding that $D_{A_{\mathrm{1g}}} \!\ll\! D_{B_{\mathrm{2g}}}$, see Fig.\ \ref{fig:Tnem-evolution}, is shown to be a consequence of the anisotropic, i.e. quasi two-dimensional
nature of these fluctuations. The analysis is performed for a model
of local spins and within a long-wavelength model of collective magnetic
fluctuations with composite order. The results of both approaches
are fully consistent with each other. The key findings of our analysis of the strain dependence of the nematic ordering temperature are summarized in Fig.\ \ref{fig:Tnem-evolution}. Anticipating the similarity in the results for the localized-spin and  the long-wavelength approach, we note the generality of the result, i.e., applying equally to systems with localized or itinerant magnetism.

\section{Strain Tuning nematic order of localized spins}

\begin{figure}[tb]
\centering
\includegraphics[width = 0.4 \textwidth]{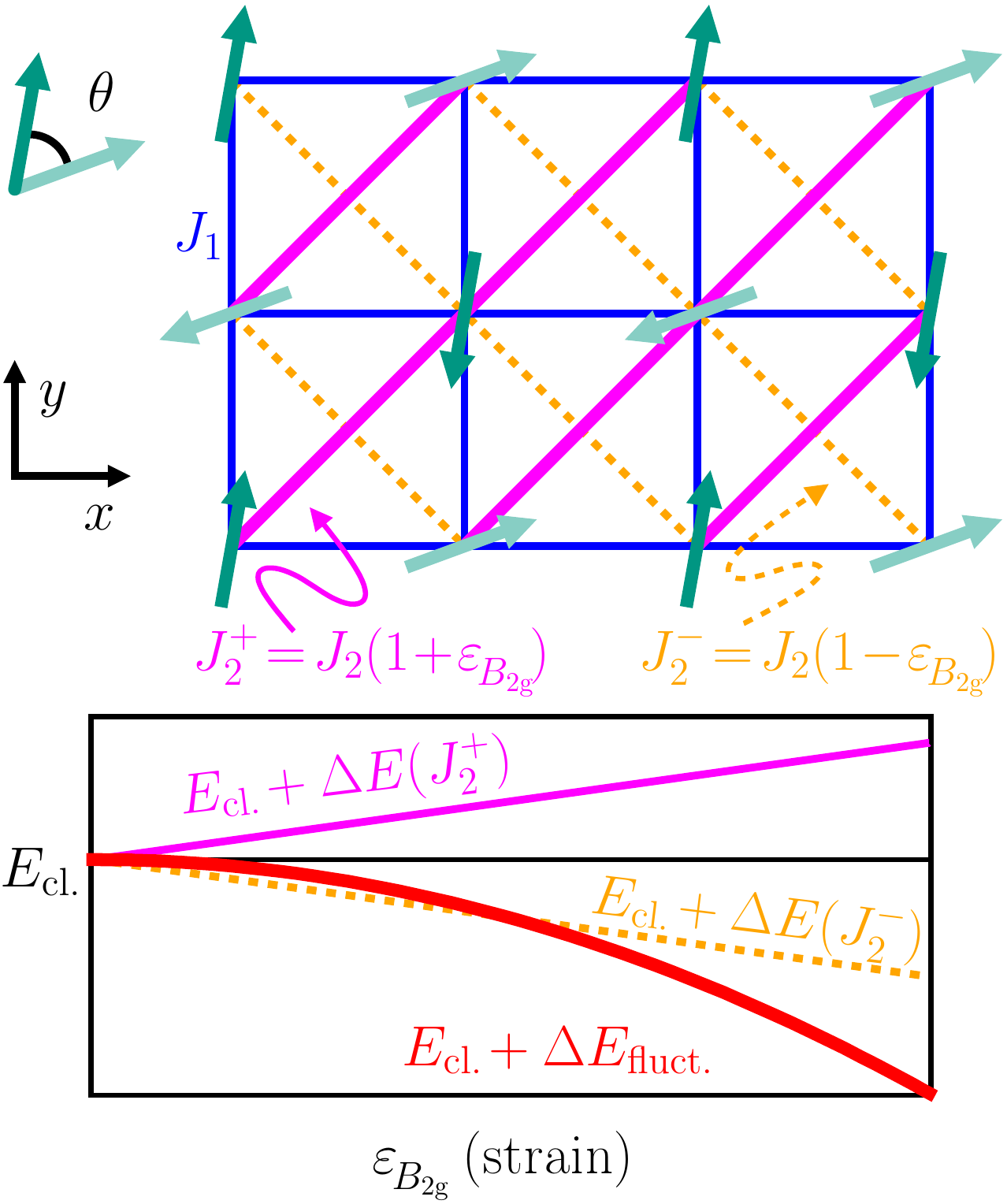}
\caption{Top: The effect of the symmetry-preserving $B_{\mathrm{2g}}$ strain can be studied in a $J_{1}^{\;}$-$J_{2}^{\pm}$ spin lattice model with asymmetric next-nearest neighbor interactions $J_{2}^{\pm} = J_{2}(1 \pm \varepsilon_{B_{\mathrm{2g}}})$. Bottom: The classical groundstate (black) is unaffected by strain, while fluctuation effects lower the energy and favor the development of a spin nematic order (green), see Eq.\ \eqref{eq:fluctuation-energy}.
}
\label{fig:lattice-model}
\end{figure}

We start our analysis with a lattice model of localized spins on a two-dimensional $J_{1}$-$J_{2}$ Heisenberg model. This model is a specific microscopic realization of the more general field theory for anisotropic three-dimensional systems that we discuss in section \ref{sec:field-theory} and offers a more microscopic insight into the strain tuning of nematic order. To be specific, we only consider the modification of the nematic transition temperature due to $\varepsilon_{B_{\mathrm{2g}}}$-strain. Without strain, the Hamiltonian is given as
\begin{align}
H=J_{1}\sum_{ \langle ij\rangle }\vec{s} (\vec{r}_{i}) \!\cdot\! \vec{s} (\vec{r}_{j})+J_{2}\sum_{ \langle \! \langle ij\rangle \! \rangle }\vec{s} (\vec{r}_{i}) \!\cdot\! \vec{s} (\vec{r}_{j}).
\end{align}
Here $\vec{s}(\vec{r}_{i})$ denotes the spin at the lattice site $\vec{r}_{i}$, $ \langle ij\rangle $ refers to a pair of nearest neighbor spins and $ \langle \! \langle ij\rangle \! \rangle$
to next-nearest neighbor spins. The important regime for us is where $J_{2}$ dominates over $J_{1}$. The $J_{1}$-$J_{2}$ Heisenberg model was shown in Ref.\ \mycite{Chandra1990} to exhibit Ising-nematic order. We closely follow the analysis of Ref.\ \mycite{Chandra1990}, yet add a external strain $\varepsilon_{B_{\mathrm{2g}}}$ that does not couple to the Ising nematic order parameter. Such strain will change the next-nearest neighbor exchange interaction according to 
\begin{align}
J_{2} \rightarrow J_{2}^{\pm} \equiv J_{2} (1\pm\kappa\varepsilon_{B_{\mathrm{2g}}}),
\end{align}
where the two signs refer to the two diagonal couplings of a square lattice, that corresponds to the single-iron unit cell description of the iron-based materials. The coupling constant $\kappa$ can be determined from a microscopic first-principles calculation. As it always as a prefactor to $\varepsilon_{B_{\mathrm{2g}}}$, we drop $\kappa$ in what follows. Fourier transformation of the Hamiltonian yields 
\begin{align}
H=\sum_{\vec{k}}J (\vec{k})\vec{s}_{\vec{k}} \!\cdot\! \vec{s}_{-\vec{k}},
\end{align}
with
\begin{align}
J (\vec{k}) &= 2J_{1} [\cos (k_{x})+\cos( k_{y})]\nonumber \\
 &\quad + 2J_{2}^{+}\cos (k_{x}+k_{y}) + 2J_{2}^{-}\cos (k_{x} - k_{y}).
\end{align}
The classical ground state energy per lattice site
\begin{align}\label{eq:classical-gs}
E_{\mathrm{cl.}} = - (J_{2}^{+} + J_{2}^{-})s^{2} =  - 2 J_{2} s^{2}.
\end{align}
of this model is unchanged by strain. The energy gain due to the enhancement of one diagonal coupling $J_{2}^{+}$ is off-set by the loss in exchange coupling along the orthogonal diagonal $J_{2}^{-}$, see Fig.\ \ref{fig:lattice-model}. This reveals that for this frustrated lattice, the effect of strain only arises due to fluctuation effects.

In what follows we will evaluate how fluctuations affect the correlation length $\xi$ and lead to a the reduction $\Delta E$ of the ground state's energy density. This program will require a separate discussion of long-wavelength and short-distance fluctuations. When combined, the energy scale $(\Delta E)\xi^{2}$ reveals the characteristic temperature where nematic order appears.

\subsection{Long-wave length fluctuations}

The long-wavelength fluctuations are described by considering $\vec{s}(\vec{k})$ for $\vec{k}$ near $\vec{Q}_{x}^{\text{2d}} \!=\! (\pi,0)$
or $\vec{Q}_{y}^{\text{2d}}\!=\! (0,\pi)$. The associated real-space modulation then reads
\begin{align}\label{eq:vec-s}
\!\! \vec{s} (\vec{r}) & \propto \vec{n}_{1} (\vec{r})\frac{e^{i\vec{Q}_{x}^{\text{2d}} \!\cdot\! \vec{r}} \!+\! e^{i\vec{Q}_{y}^{\text{2d}} \!\cdot\! \vec{r}}}{2}
+ \vec{n}_{2} (\vec{r})\frac{e^{i\vec{Q}_{x}^{\text{2d}} \!\cdot\! \vec{r}} \!-\! e^{i\vec{Q}_{y}^{\text{2d}} \!\cdot\! \vec{r}}}{2},
\end{align}
where $\vec{n}_{a} (\vec{r})^{2}=1$ are unit vectors and the coordinate $\vec{r} = (x,y)$ becomes a continuous variable. The two coupled, interpenetrating N\'eel sub-lattices magnetizations $\vec{n}_{1,2}$ can be modeled in a nonlinear sigma model \mycite{Chandra1990},
\begin{align}
\!\!
S \!&=\! \frac{1}{2g} \!\int\!\! d\vec{r}\, \Big\{\! \Big[\!\!\sum_{j=1,2}\!\! (\nabla\vec{n}_{a})^{2} \Big]
 \!+ 2 \alpha (\partial_{x}\vec{n}_{1} \!\cdot\! \partial_{x}\vec{n}_{2} \!-\! \partial_{y}\vec{n}_{1} \!\cdot\! \partial_{y}\vec{n}_{2}) \nonumber \\
 \label{eq:nlsm}
 &\qquad \qquad +  2 \varepsilon_{B_{\mathrm{2g}}} (\partial_{x}\vec{n}_{1} \!\cdot\! \partial_{y}\vec{n}_{1}+\partial_{x}\vec{n}_{2} \!\cdot\! \partial_{y}\vec{n}_{2})\Big\}.
\end{align}
Here $g=T/2J_{2}s^{2}$ is the stiffness and $\alpha=J_{1}/J_{2}$. One can eliminate the strain term proportional to $\varepsilon_{B_{\mathrm{2g}}}$ through two successive coordinate transformations. First, we introduce $z_{\mp}=(x \mp y)/[2(1 \mp \varepsilon_{B_{\mathrm{2g}}})]^{1/2}$, i.e.\ a rotation by $\pi/4$ with a simultaneous stretching/compression along the two directions. The second transformation $(\tilde{x}, \tilde{y}) = (z_{+} + z_{-}, z_{+} - z_{-})/\sqrt{2}$ rotates the coordinates back (by $-\pi/4$) (without compressing or stretching). The action then takes the form
\begin{align}\label{eq:nlsigma transformed}
\!
S \!=\!\frac{1}{2\tilde{g}} \!\int\!\! d\tilde{\vec{r}}\, \Big\{\! \Big[\!\!\sum_{j=1,2}\!\! (\tilde{\nabla}\vec{n}_{a})^{2} \Big]
 \!+ 2 \tilde{\alpha} (\partial_{\tilde{x}}\vec{n}_{1} \!\cdot\! \partial_{\tilde{x}}\vec{n}_{2} \!-\! \partial_{\tilde{y}}\vec{n}_{1} \!\cdot\! \partial_{\tilde{y}}\vec{n}_{2}),
\end{align}
with $\tilde{g} = g/(1-\varepsilon_{B_{\mathrm{2g}}}^{2})^{1/2}$ and $\tilde{\alpha} = \alpha/(1-\varepsilon_{B_{\mathrm{2g}}}^{2})^{1/2}$. This is precisely the model without strain, yet with a reduced (effective) next nearest neighbor interaction
\begin{align}
J_{2} \rightarrow J_{2}(1-\varepsilon_{B_{\mathrm{2g}}}^{2})^{1/2}<J_{2}.
\end{align}
It is important to note that fluctuations are affected by strain while the classical ground-state energy is not. This gives rise to a shortening of the magnetic correlation length $\xi$. For the two-dimensional spin model follows from the usual renormalization group procedure \mycite{Chandra1990} that
\begin{align}\label{eq:correlation-length}
\!\!
\xi \sim a_{0} e^{2 \pi / \tilde{z} \tilde{g}} \approx a_{0} \exp\Big\{\frac{2 \pi}{z g} [ 1 \!-\!  (1+\alpha/4)\varepsilon_{B_{\mathrm{2g}}}^2/2 )] \Big\},
\end{align}
with $a_{0}$ of the order of the lattice constant, $\tilde{z} \!=\! z(\tilde{\alpha})$ and $z = z(\alpha) \equiv 2 \alpha/ (\arcsin(\alpha) + \alpha \sqrt{1-\alpha})$. The strain-induced reduction of the spin-wave stiffness is the is the predominant effect of strain on long wavelength magnetic fluctuations. 

In performing the above coordinate transformations one has to be careful as they may change the boundary conditions and thus the symmetry of the system. This turns out to be a problem if one considers $\varepsilon_{B_{\mathrm{1g}}}$ strain that couples to the Ising-nematic order parameter. For $\varepsilon_{B_{\mathrm{2g}}}$ discussed here this problem does not exist.

\subsection{Short-distance fluctuations}

Whereas $B_{\mathrm{2g}}$ strain modifies the stiffness of long-wave fluctuations, is will affect the system's energy through short distance fluctuations. This energy depends on the angle $\theta \equiv \arccos(\vec{n}_{1\!}  \cdot  \vec{n}_{2})$ between the two sub-lattices, see Fig.\ \ref{fig:lattice-model}.
To quantify the effect of strain we perform a $1/s$ spin-wave analysis of the Heisenberg model (similar to Ref.\ \mycite{Chandra1990}) with distinct strain-induced exchange interactions along the two diagonals. The spin-wave spectrum of this problems is given as
\begin{align}
  \omega (\vec{k})=4 s J_{2}\sqrt{A (\vec{k})^{2}-B (\vec{k})^{2}},
\end{align}
where 
\begin{align}
A (\vec{k}) &= 1 + \alpha [\cos^{2}(\theta)\cos k_{x}+\sin^{2}(\theta)\cos k_{y}], \\
B (\vec{k}) &= b (\vec{k})+\alpha [\cos^{2}(\theta)\cos k_{y} + \sin^{2}(\theta)\cos k_{x}],\quad
\end{align}
and 
\begin{align}
b (\vec{k})=\frac{\alpha}{2}\sum_{\sigma = \pm} (1+ \sigma \varepsilon_{B_{\mathrm{2g}}})\cos (k_{x} + \sigma k_{y}).
\end{align}
This allows us to analyze the free energy corrections
\begin{align}
\Delta F = T\sum_{\vec{k}}\ln \Big[ \sinh \Big(\frac{\omega (\vec{k})}{2T}\Big)\Big],
\end{align}
due to spin-wave excitations, which---in the limit $T \!\to\! 0$---corresponds to the correction to the ground state energy $\Delta E=\Delta F (T=0)$. Performing the momentum integration, we obtain the additional biquadratic
exchange energy
\begin{align}\label{eq:fluctuation-energy}
\Delta E=\gamma_{Q}\frac{J_{2}s}{2}\alpha^{2} [1+ (\vec{n}_{1} \!\cdot\! \vec{n}_{2})^{2}],
\end{align}
where 
\begin{align}\label{eq:gamma-Q}
\gamma_{Q}=\gamma_{Q}^{ (0)} (1+ \zeta \varepsilon_{B_{\mathrm{1g}}}^{2}).
\end{align}
The coefficient $\gamma_{Q}^{ (0)}\approx0.26025$ describes the situation without strain and was already given in Ref.\ \mycite{Chandra1990}. For our considerations it is more important to determine the change in the bi-quadratic interaction due to strain which is characterized by the coefficient $\zeta \approx 0.17273 > 0$. Thus, the biquadratic spin interaction \textit{increases} due to strain. 

\subsection{The nematic transition temperature}

Combining long and short wave-length excitations finally allows us to determine the nematic transition temperature $T_{\mathrm{nem}}.$ The order parameter of the nematic state is the Ising variable%
\begin{align}
  \phi= \langle \vec{n}_{1} \!\cdot\! \vec{n}_{2}\rangle,
\end{align}
which relates to the angle $\theta$ between the two sublattices via $\cos(\theta) = \phi$. To determine $T_{\mathrm{nem}}$ we consider the typical interaction energy of a region of size $\xi^{2}$. If this energy is comparable to the temperature, one expects the nematic phase transition to take place \mycite{Chandra1990}. This gives rise to the criterion
\begin{align}
\Big[\frac{\xi (T_{\mathrm{nem}})}{a_{0}}\Big]^{2} \Delta E &= k_{B}T_{\mathrm{nem}}
\end{align}
We obtain two opposite trends due to external strain: On the one hand, the correlation length gets smaller which reduces the transition temperature. On the other hand, $\Delta E$ gets larger, which tends to enhance $T_{\mathrm{nem}}$. For any system that is near a second-order magnetic phase transition the correlation length is about to diverge. Even for weak first-order transitions, the correlation length above the magnetic ordering temperature is exponentially large, see Eq.\ \eqref{eq:correlation-length}. As a result, the strain modification of the correlation length is always the dominant one. Thus, we find that within the spin-induced nematic theory, $B_{\mathrm{2g}}$ strain clearly decreases the nematic transition temperature, in agreement with the experimental observation \mycite{Ikeda2018}. This suppression of $T_{\mathrm{nem}}$ is quadratic in $\varepsilon_{B_{\mathrm{2g}}}$ as suggested by the expressions \eqref{eq:correlation-length} and \eqref{eq:gamma-Q}.

Since the analysis of this section was performed for a two-dimensional system with exponentially growing magnetic correlation length, it is important to analyze the role of three-dimensional, albeit anisotropic spin correlations, relevant for many iron-based materials. This analysis will be performed in the next section. 

\section{Strain tuning for spin-induced vestigial order}\label{sec:field-theory}

In this section, we consider the long-wavelength theory of collective magnetic degrees of freedom along the lines of Ref.\ \mycite{Fernandes2012a}. As mentioned above, this approach may equally apply to the collective response of localized spins, or to magnetism of itinerant moments. We consider a generic spin configuration
\begin{align}
\vec{\vec{s}} (\vec{R})=\vec{m}_{x} (\vec{R})e^{i\vec{Q}_{x} \!\cdot \vec{R}}+\vec{m}_{y} (\vec{R})e^{i\vec{Q}_{y} \!\cdot \vec{R}},
\end{align}
where the $\vec{m}_{x,y} (\vec{R})$ vary smoothly in space. In distinction to Eq.\ \eqref{eq:vec-s} the fields $\vec{m}_{a}$ are not unit vectors. As we are analyzing anisotropic three-dimensional systems, the ordering vectors are now given as $\vec{Q}_{x\!} \!=\! (\pi,0,0)$ and $\vec{Q}_{y\!} \!=\! (0,\pi,0)$.\ We perform a continuum's description for the coordinates within the planes, but keep the discrete
lattice structure for the third dimension, with layer index $l$. Thus, we express the three-dimensional coordinates $\vec{R}= (\vec{r},la_{z})$ in terms of the two-dimensional vector $\vec{r}= (x,y)$ and the discrete layer index $l$. Following Ref.\ \mycite{Fernandes2018-arXiv} we combine the two vectors into 
%
$
\vec{m}= (\vec{m}_{x},\vec{m}_{y}).
$
%
and obtain the effective action of the problem
\begin{align}\label{eq:action}
S=\sum_{l}\int d\vec{r}\mathcal{L} (\vec{m},\partial_{\beta}\vec{m})
\end{align}
where the Lagrangian $\mathcal{L}\!=\!\mathcal{L}_{\parallel}+\mathcal{L}_{\perp}$
consists of an intra-layer term
\begin{align}
\mathcal{L}_{\mathrm{\parallel}} & = \frac{r_{0}}{2}\vec{m}_{l}\tau_{0}\vec{m}_{l}+\frac{u}{4} (\vec{m}_{l}\tau_{0}\vec{m}_{l})^{2} \\
\nonumber
 &\quad - \frac{g}{4} (\vec{m}_{l}\tau_{3}\vec{m}_{l})^{2}+\frac{v}{4} (\vec{m}_{l}\tau_{1}\vec{m}_{l})^{2}+\frac{1}{2}\partial_{\beta}\vec{m}_{l}\tau_{0}\partial_{\beta}\vec{\vec{m}}_{l} \\
 \nonumber
 &\quad + \frac{\alpha}{2} (\partial_{x}\vec{m}_{l}\tau_{3}\partial_{x}\vec{\vec{m}}_{l}-\partial_{y}\vec{m}_{l}\tau_{3}\partial_{y}\vec{\vec{m}}_{l})
\end{align}
and a coupling between nearest neighboring layers 
\begin{align}\label{eq:Lagrange-interlayer}
\mathcal{L}_{\mathrm{\perp}}=q_{0}^{2}\vec{m}_{l}\tau_{0}\vec{m}_{l+1}.
\end{align}
The Pauli matrices $\tau_{\alpha}$ act in the space of two ordering vectors, e.g.\ $\vec{m}\tau_{3}\vec{m} \!=\! \vec{m}_{x} \!\cdot\! \vec{m}_{x} \!-\! \vec{m}_{y} \!\cdot\! \vec{m}_{y}$. Depending on the sign and the magnitude of the coupling constants $u$, $g$ and $v$, several magnetic phases and their associated vestigial orders have been discussed \mycite{Fernandes2016}: These are the stripe antiferromagnetic ($\vec{m}\tau_{3}\vec{m}$), the charge-spin density-wave ($\vec{m}\tau_{1}\vec{m}$). A third nematic order $\!\propto\! \vec{m}_{1}\!\times\!\vec{m}_{2}$ (not considered here), associated with a spin-vortex crystal has been identified. Microscopic expressions for the phenomenological parameters are given elsewhere, see Ref.\ \mycite{Fernandes2012a}. Note that our analysis assumes non-critical responses of these phenomenological parameters to strain. More specifically their strain-dependence is neglected in the following. Tuning of nematicity is then caused by the modification of critical fluctuations at finite strain.

In the presence of strain $\varepsilon_{\alpha\beta}$ the response of collective spin modes is governed by a modified action
\begin{align}\label{eq:action-epsilon}
S_{\varepsilon}=S-\sum_{l}\int d\vec{r}\sum_{\alpha\beta}\varepsilon_{\alpha\beta}T_{\alpha\beta}.
\end{align}
with the stress tensor
\begin{align}
T_{\alpha\beta}=\frac{\partial\mathcal{L}}{\partial (\partial_{\alpha}\vec{m})} \!\cdot\! \partial_{\beta}\vec{m}-\delta_{\alpha\beta}\mathcal{L}.
\end{align}
In our subsequent analysis we will work at constant strain. This is clearly adequate to describe measurements such as the elasto-resistivity \mycite{Chu2012}. On the other hand, experiments on unstrained samples should rather be performed at fixed stress. For the description of Ising nematic order this requires to include harmonic elastic terms, characterized by bare elastic constants. As shown in Ref.\ \mycite{Fernandes2010} this gives rise to an enhancement of the nematic coupling constant, hence enhancing the nematic transition temperature compared to the value of a purely electronic system. For the experiments of Ref.\ \mycite{Ikeda2018} one has to keep in mind that fixed strain $\varepsilon_{B_{\mathrm{2g}}} \neq 0$ still corresponds to fixed stress in the other symmetry channels. Thus, the above renormalizations of the nematic coupling due to fluctuations of $\varepsilon_{B_{\mathrm{1g}}}$ should nevertheless be included. We will expand on these issues below, when we make contact to experiment. 

In the following, we focus on in-plane strain; in particular we introduce the strain combinations 
\begin{align}
\varepsilon_{A_{\mathrm{1g}}} & \equiv (\varepsilon_{xx}+\varepsilon_{yy})/2, \\
\varepsilon_{B_{\mathrm{1g}}} & \equiv (\varepsilon_{xx}-\varepsilon_{yy})/2, \\
\varepsilon_{B_{\mathrm{2g}}} & \equiv (\varepsilon_{xy}+\varepsilon_{yx})/2,
\end{align}
and similar for the stress tensor. In this notation the strain-stress term in Eq.\ \eqref{eq:action-epsilon} takes the form
\begin{align}
\label{eq:strain-effect}
S_{\varepsilon} \!-\! S &= - 2 \sum_{l}\int \! d\vec{r} (\varepsilon_{A_{\mathrm{1g}}} T_{A_{\mathrm{1g}}} \!+\! \varepsilon_{B_{\mathrm{1g}}} T_{B_{\mathrm{1g}}} \!+\! \varepsilon_{B_{\mathrm{2g}}} T_{B_{\mathrm{2g}}}).
\end{align}
For the trivially transforming $A_{\mathrm{1g}}$ combination follows
\begin{align}
T_{A_{\mathrm{1g}}} &=
\frac{1}{2}\partial_{\alpha}\vec{m}\tau_{0}\partial_{\alpha}\vec{m}
+ \frac{\alpha}{2} (\partial_{x}\vec{m}\tau_{3}\partial_{x}\vec{m}\nonumber \\
 &\quad - \partial_{y}\vec{m}\tau_{3}\partial_{y}\vec{m}) -\mathcal{L},
\end{align}
while the stress associated to the two nontrivial irreducible representations $B_{\mathrm{1g}}$ and $B_{\mathrm{2g}}$ read 
\begin{align}
T_{B_{\mathrm{1g}}} & = \partial_{x}\vec{m}\tau_{0}\partial_{x}\vec{m}-\partial_{y}\vec{m}\tau_{0}\partial_{y}\vec{m} \\ \nonumber
 &\quad+ \alpha (\partial_{x}\vec{m}\tau_{3}\partial_{x}\vec{m}+\partial_{y}\vec{m}\tau_{3}\partial_{y}\vec{m}), \\
T_{B_{\mathrm{2g}}} & = \partial_{x}\vec{m}\tau_{0}\partial_{y}\vec{m}.
\end{align}
The combination $(T_{xy}-T_{yx}) / 2 = \alpha \partial_{y} \vec{m}\tau_{3}\partial_{x}\vec{m}$ describes the response to a rotation about the $z$ axis and will not be discussed further. 

In our treatment of anisotropic three-dimensional systems, the dispersion in the third direction is characterized by the coupling between neighboring layers of Eq.\ \eqref{eq:Lagrange-interlayer}. To simplify our notation we use $\vec{q}^{2}=q_{x}^{2}+q_{y}^{2}+q_{z}^{2}$ where $q_{z}^{2}$ stands in fact for $2q_{0}^{2}[1-\cos(q_{z}/q_{0})]$. Here $q_{0}$ is a measure of the in-plane versus out-of-plane anisotropy. The 2d case is obtained in the limit $q_{0} \!\to\! 0$, while for $q_{0} \!\to\! \infty$ the isotropic 3d case is recovered. By parametrizing the in-plane momentum vector as $(q_{x},q_{y})=q (\cos\varphi,\sin\varphi)]$ (consequently $\vec{q}^{2} = q^{2} + q_{z}^{2}$), and after introducing an ulta-violett cut-off $q<\Lambda$, the momentum integral of a function $f(\vec{q})$ takes the form 
\begin{align}
\int_{\vec{q}}f(\vec{q})=\int\limits _{-\pi q_{0}}^{\pi q_{0}}\frac{dq_{z}}{2\pi}\int\limits _{0}^{\Lambda}\frac{dq\ q}{2\pi}\int\limits _{0}^{2\pi}\frac{d\varphi}{2\pi}f(\vec{q}).\label{eq:convention}
\end{align}

\subsection{Analysis without strain}

The subsequent analysis extends the approach used in Ref.\ \mycite{Fernandes2012a} to the case of finite strain. For completeness, we briefly summarize the zero-strain case. After introducing the Hubbard-Stratonovich fields $\eta$ and $\phi$ that are conjugate to $\vec{m}\tau_{0}\vec{m}$ and $\vec{m}\tau_{3}\vec{m}$ respectively, we can integrate out the quadratic action for the fields $\vec{m}$ and obtain an effective action for the conjugate fields 
\begin{align}
S=\frac{N}{2} \! \sum_{l} \!\! \int \!\! d^{2}\vec{r} \Big[\frac{3u}{2}\eta^{2} \!+\! \frac{3g}{2} \phi^{2} \Big] \!-\! \mathrm{tr} \ln(G_{1}^{-1}) \!-\! \mathrm{tr}\ln(G_{2}^{-1}).
\end{align}
where the Green's functions associated with $\vec{m}_{1,2}$ read
\begin{align}
G_{1,2}^{-1}(\vec{r}, l; \vec{r}', l') &= \delta(\vec{r} \!-\! \vec{r}') \delta_{ll'} \big[r_{0}-i3u\eta \mp 3 g \phi\label{eq:inv-G} \\ \nonumber
 &\qquad \qquad \qquad \quad +  \nabla_{\vec{r}'}^{2}\mp\alpha(\partial_{x'}^{2}-\partial_{y'}^{2})\big] \\ \nonumber
 &\quad +  \delta (\vec{r} \!-\! \vec{r}') (\delta_{l,l' + 1} + \delta_{l,l'- 1}) q_{0}^{2}. 
\end{align}
In the large-$N$ limit, the partition function may be evaluated at the saddle-point solution. Then 
\begin{align}
\phi= \langle \vec{m}\tau_{z}\vec{m}\rangle = \langle \vec{m}_{x}^{2}-\vec{m}_{y}^{2}\rangle 
\end{align}
emerges as the Ising nematic order parameter. 

Expressing the saddle-point equations in Fourier space, we assume here that $\eta$ and $\phi$ are not coordinate-dependent, we obtain the self-consistency equations 
\begin{align}
r &=r_{0}+6u\int_{\vec{q}}\frac{r+\vec{q}^{2}}{(r+\vec{q}^{2})^{2}-[3g\phi-\alpha(q_{x}^{2}-q_{y}^{2})]^{2}}\label{eq:self-consistency-eq-3} \\
\phi &= \int_{\vec{q}}\frac{6g\phi-2\alpha(q_{x}^{2}-q_{y}^{2})}{(r+\vec{q}^{2})^{2}-[3g\phi-\alpha(q_{x}^{2}-q_{y}^{2})]^{2}},\label{eq:self-consistency-eq-4}
\end{align}
where we introduced $r=r_{0}-i3u\eta$ and used the convention of
Eq.\ \ref{eq:convention}.

In order to highlight solution of this set of equations, let us consider the case $\alpha=0$ first. The more general case $\alpha\neq0$ will be treated below when we include finite strain. The renormalization of $r_{0} \!\to\!  r$ gives rise to a change in the in-plane magnetic correlation length $\xi$, via
\begin{align}
r - 3g\phi=\xi^{-2}
\end{align}
In absence of nematic order $\phi \!=\! 0$ we use $\bar{r} \!\equiv\! r (\phi=0)$ which obeys 
\begin{align}
\bar{r} & =r_{0}+6uI_{1}(\bar{r}).\label{eq:self-consistent}
\end{align}
where $I_{1}(\bar{r})$ belongs to a series of reappearing integrals $I_{n}(\bar{r})$, for which we introduce a unified notation, see Appendix \ref{app:integrals}. Treating the onset of nematicity perturbatively we write $r \!\approx\! \bar{r} + d\phi^{2}$ and find $d \!=\! (3g)^{2}6uI_{3}(\bar{r}) / [1+6uI_{2}(\bar{r})]$ from Eq.\ \eqref{eq:self-consistency-eq-3}. Similarly, Eq.\ \eqref{eq:self-consistency-eq-4} takes the form
\begin{align}
\!\!\!
0 &= a \phi + b\phi^{3} \;\;\text{with} \;\;
\left\{
\begin{aligned}
a &=1-6gI_{2}(\bar{r}) \\
b &= 6g[2 d I_{3}(\bar{r})-(3g)^{2}I_{4}(\bar{r})].
\end{aligned}
\right.\!\!\!
\label{eq:effektiv-phi-hoch4}
\end{align}
This is the equation of state for $\phi$ that can be interpreted
as being due to an effective $\phi^{4}$-theory,
\begin{align}\label{}
   F_{\mathrm{nem}} = \frac{a}{2}\phi^{2} + \frac{b}{4} \phi^{4}
\end{align}
for the nematic order parameter: At high temperature, when the coefficient $1-6gI_{2}(\bar{r})$ is positive, no nematic order exists. A vanishing of this coefficient defines the nematic transition temperature $T_{\mathrm{nem}}$ via the temperature dependence of $\bar{r}$. Substituting $I_{2}(\bar{r})$ by its expression in an anisotropic 3d system, as derived in Appendix \ref{app:anis-3d}, the nematic phase sets in when $\sqrt{\bar{r}(\bar{r}+4q_{0}^{2})}=3gq_{0}/2\pi$. The sign of the square bracket in Eq.\ \eqref{eq:effektiv-phi-hoch4} at the onset of nematicity [proportional to $3g(u-2g)-16\pi^{2}q_{0}^{2}$ in the anisotropic 3d system] discriminates between a first (negative) and second order (positive) phase transition. One readily sees that an isotropic three dimensional system always undergoes a first order nematic transition, while split second order transitions occur in sufficiently anisotropic systems, see also Ref.\ \mycite{Fernandes2012a}.

\subsection{Strain tuning Ising nematic order}

\subsubsection{$A_{\mathrm{1g}}$-strain}
When applying an external strain in the $A_{\mathrm{1g}}$-channel,
$\varepsilon_{xx} \!=\! \varepsilon_{yy} \!=\! \varepsilon_{A_{\mathrm{1g}}}$ the inverse Green's
functions \eqref{eq:inv-G} are modified to 
\begin{align}
G_{1,2}^{-1}(\vec{r}, l; \vec{r}', l') &= \delta(\vec{r} \!-\! \vec{r}') \delta_{ll'} \big[\gamma r_{0} \!-\!  i3\gamma u\eta\mp3\gamma g\phi-\gamma\nabla_{\!\vec{r}'}^{2} \nonumber \\
 &\qquad \quad + 2\varepsilon_{A_{\mathrm{1g}}} (\partial_{x'}^{2}+\partial_{y'}^{2})\mp\alpha(\partial_{x'}^{2}-\partial_{y'}^{2})\big] \nonumber \\ 
 &\quad + \delta (\vec{r} \!-\! \vec{r}') (\delta_{l,l'+1} + \delta_{l,l'-1})q_{0}^{2}
\end{align}
with $\gamma=(1+2\varepsilon_{A_{\mathrm{1g}}})$ renormalizing the bare parameters $r_{0}$, $u$, and $g$ as well as the gradient term. From this renormalization follows immediately, that $\varepsilon_{A_{\mathrm{1g}}}$ affects the transition temperatures to linear order. The self-consistency equations for $r$ and $\phi$ now read 
\begin{align}
r & = r_{0} + \int_{\vec{q}}\frac{6u[\gamma(r+\vec{q}^{2})-2\varepsilon_{A_{\mathrm{1g}}} q^{2}]}{\mathcal{D}_{A_{\mathrm{1g}}} (\vec{q},\varepsilon_{A_{\mathrm{1g}}})}\label{eq:self-consistency-eq-7}\\
\phi & =\int_{\vec{q}}\frac{6\gamma g\phi-2\alpha(q_{x}^{2}-q_{y}^{2})]}{\mathcal{D}_{A_{\mathrm{1g}}} (\vec{q},\varepsilon_{A_{\mathrm{1g}}})}\label{eq:self-consistency-eq-8}
\end{align}
with the integrand's denominator
\begin{align}
\mathcal{D}_{A_{\mathrm{1g}}} (\vec{q},\varepsilon_{A_{\mathrm{1g}}}) &= [\gamma(r+\vec{q}^{2})-2\varepsilon_{A_{\mathrm{1g}}} q^{2}]^{2}  \\ \nonumber
&\quad - [3\gamma g\phi-\alpha(q_{x}^{2}-q_{y}^{2})]^{2}.
\end{align}
Considering the effects of $\alpha$ and $\varepsilon_{A_{\mathrm{1g}}}$ as perturbations
to the system, motivates the Ansatz 
\begin{align}
r\approx\bar{r}+d\phi^{2}+d'\alpha^{2}+f_{A_{\mathrm{1g}}}\varepsilon_{A_{\mathrm{1g}}}.
\end{align}
Specifically, the above self-consistency equations respond linearly to an external strain while the combination $q_{x}^{2}-q_{y}^{2}$ only allows for even powers in $\alpha$. By expanding the first equation and equating coefficients we find $d$ as in the case without strain as well as
\begin{align}
d' & =\frac{6uJ_{3}^{2,0}(\bar{r})}{1+6uI_{2}(\bar{r})}, & f_{A_{\mathrm{1g}}} & =\frac{12u[J_{2}^{0,1}(\bar{r})-I_{1}(\bar{r})]}{1+6uI_{2}(\bar{r})},\label{-1}
\end{align}
with the functions $J_{n}^{{\ell,m}}(\bar{r})$ defined in Appendix \ref{app:integrals}. Notice, if the index $n$ subscript exceeds the sum of the superscripts $\ell+m$ by more than 1, the integral is convergent for the anisotropic 3d case. For $n \!=\! \ell \!+\! m \!+\! 1$ the integral is logarithmically divergent. Expanding the second self-consistency-equation \eqref{eq:self-consistency-eq-8} in the small parameters $\psi$, $\alpha$ and $\varepsilon_{A_{\mathrm{1g}}}$ yields the equation of state $0 \!=\! a_{A_{\mathrm{1g}}}\!(\varepsilon_{A_{\mathrm{1g}}})\phi + b\phi^{3}$, see Eq.\ \eqref{eq:effektiv-phi-hoch4}, with the linear coefficient
\begin{align}\label{eq:GL-A1g}
a_{A_{\mathrm{1g}}}\!(\varepsilon_{A_{\mathrm{1g}}}) &= a + 6g\alpha^{2}[2d'I_{3}(\bar{r})-3J_{4}^{2,0}(\bar{r})] \\
 &\quad + 12g\varepsilon_{A_{\mathrm{1g}}}[f_{A_{\mathrm{1g}}}I_{3}(\bar{r})+I_{2}(\bar{r})-2J_{3}^{0,1}(\bar{r})]\nonumber \\
 &\quad + 6g\varepsilon_{A_{\mathrm{1g}}}^{2}[3f_{A_{\mathrm{1g}}}^{2}I_{4}(\bar{r})+4f_{A_{\mathrm{1g}}}I_{3}(\bar{r})+4I_{2}(\bar{r})\nonumber \\
 &\qquad- 12f_{A_{\mathrm{1g}}}J_{4}^{0,1}(\bar{r})-16J_{3}^{0,1}(\bar{r})+12J_{4}^{0,2}(\bar{r})].\nonumber
\end{align}
The coefficient $a_{A_{\mathrm{1g}}}\!(\varepsilon_{A_{\mathrm{1g}}})$ now determines the ground state at finite strain and, in particular, the onset of the nematic order. As anticipated the $\mathrm{A}_{\mathrm{1g}}$-strain changes \emph{linearly} the onset of nematic order, as featured by a term $\propto\! \varepsilon_{A_{\mathrm{1g}}}$ in the linear coefficient $a$ of the effective $\phi^{4}$ theory.

Anticipating a discussion with the results obtained in other strain channels, the expansion in $\varepsilon_{A_{\mathrm{1g}}}$ has been carried to quadratic order. In this context it is important to note that the coefficient $f_{A_{\mathrm{1g}}}$ and the square bracket determining the quadratic correction $\propto\! \varepsilon_{A_{\mathrm{1g}}}^{2}\!$ remain bounded for strongly anisotropic systems. As we shall see, this is in contrast to the quadratic terms in the $\mathrm{B}_{\mathrm{1g}}$ and $\mathrm{B}_{\mathrm{2g}}$ channels where the coefficient $\propto\! \varepsilon_{B_{\mathrm{2g}}}^{2}\!$ logarithmically diverges upon approaching the two-dimensional limit.

\subsubsection{$B_{\mathrm{1g}}$-strain}

Next, we consider an external strain with the components $\varepsilon_{xx}\!=\!-\varepsilon_{yy}\!=\!\varepsilon_{B_{\mathrm{1g}}}$. Following the derivation scheme discussed above, the layer-diagonal part of inverse Green's functions \eqref{eq:inv-G} will be augmented by additional terms while the inter-layer terms remain unchanged
\begin{align}
G_{1,2}^{-1}(\vec{r}, l; \vec{r}', l) & = \delta(\vec{r} \!-\! \vec{r}')\big[r_{0}-i3u\eta\mp3g\phi\\
\nonumber
 &\quad- \nabla_{\vec{r}'}^{2}\mp\alpha(\partial_{x'}^{2}-\partial_{y'}^{2}) \\
\nonumber
 &\quad+ 2\varepsilon_{B_{\mathrm{1g}}}(\partial_{x'}^{2}-\partial_{y'}^{2}) \pm 2 \varepsilon_{B_{\mathrm{1g}}} \alpha(\partial_{x'}^{2}+\partial_{y'}^{2})\big]
\end{align}
The self-consistency equations for $r$ and $\psi$ now read
\begin{align}\label{eq:self-consistency-B1g-1}
r & = r_{0} + \int_{\vec{q}}\frac{6u[r+\vec{q}^{2}-2\varepsilon_{B_{\mathrm{1g}}} (q_{x}^{2}-q_{y}^{2})]}{\mathcal{D}_{B_{\mathrm{1g}}} (\vec{q},\varepsilon_{B_{\mathrm{1g}}})}, \\
\label{eq:self-consistency-B1g-2}
\phi & = \int_{\vec{q}}\frac{6g\phi-2\alpha(q_{x}^{2}-q_{y}^{2})+4\alpha \varepsilon_{B_{\mathrm{1g}}}  q^{2}}{\mathcal{D}_{B_{\mathrm{1g}}} (\vec{q},\varepsilon_{B_{\mathrm{1g}}})},
\end{align}
with 
\begin{align}
\mathcal{D}_{B_{\mathrm{1g}}} (\vec{q},\varepsilon_{B_{\mathrm{1g}}}) & =  [r+\vec{q}^{2}-2\varepsilon_{B_{\mathrm{1g}}} (q_{x}^{2}-q_{y}^{2})]^{2} \\ \nonumber
 &\,\quad - [3g\phi-\alpha(q_{x}^{2}-q_{y}^{2})+2\alpha\varepsilon_{B_{\mathrm{1g}}} q^{2}]^{2}.
\end{align}
In the equation for $\phi$, the term $\propto\!\alpha \varepsilon_{B_{\mathrm{1g}}}$ of the integrand's numerator implies that the nematic order exists at any temperature. In fact, this term generically produces a finite value for $\phi$. As shall become clear below, the symmetry-breaking term acts as an Ising field $h_{\mathrm{eff}}$ lifting the phase transition. An expansion of Eq.\ \eqref{eq:self-consistency-B1g-1} using the Ansatz $r\approx\bar{r}+d\phi^{2}+d'\alpha^{2}+f_{B_{\mathrm{1g}}}\varepsilon_{B_{\mathrm{1g}}}^{2}$ yields the known values for $d$ and $d'$ as well as $f_{B_{\mathrm{1g}}}=4d'$. Expanding the equation of state \eqref{eq:self-consistency-B1g-2} for $\phi$ provides 
\begin{align}\label{eq:GL-B1g}
   h_{\mathrm{eff}} = a \phi + b \phi^{3}
\end{align}
with effective source field  $h_{\mathrm{eff}} \!=\! 4\alpha\varepsilon_{B_{\mathrm{1g}}}[J_{2}^{0,1}(\bar{r})-2J_{3}^{2,0}(\bar{r})]$ induced by strain. Since $\phi$ and $\varepsilon_{B_{\mathrm{1g}}}$ transform according to the same irreducible representation, it is natural to expect the nemato-elastic term $\lambda_{\mathrm{n.e.}}\phi \varepsilon_{B_{\mathrm{1g}}}$ in the free energy expansion. Our analysis shows that the nemato-elastic coupling constant is 
\begin{equation}
\lambda_{\mathrm{n.e.}}=4\alpha [J_{2}^{0,1}(\bar{r})-2J_{3}^{2,0}(\bar{r})],
\end{equation}
and simplifies to $\lambda_{\mathrm{n.e.}} = \alpha q_{0}/2\pi$ in the 2d limit. Subleading corrections in $\varepsilon_{B_{\mathrm{1g}}}$ or $\alpha$ to the Landau parameters $a$ and $b$ have been omitted here%
%
\footnote{Quadratic corrections in $\alpha$ and $\varepsilon_{B_{\mathrm{1g}}}$ on the left-hand side of Eq.\ \eqref{eq:GL-B1g} turn out to be identical to the corrections in Eq.\ \eqref{eq:GL-B2g}. Only in the accidental case where either $\varepsilon_{B_{\mathrm{1g}}}$ or $\alpha$ identically vanishes, the nematic transition survives and is merely shifted by a term proportional to $\alpha^{2}$ or $\varepsilon_{B_{\mathrm{1g}}}^{2}$ respectively.}.
The parameters $\alpha$ and $\varepsilon_{B_{\mathrm{1g}}}$ now appear to linear order
on the left-hand side of the above equation and generically induce
a finite nematic order at any temperature. The nematic order is 
\begin{align}\label{eq:induced order parameter}
\phi \approx \frac{\lambda_{\mathrm{n.e.}}}{a} \varepsilon_{B_{\mathrm{1g}}} =  4 \alpha \frac{J_{2}^{0,1}(\bar{r})-2J_{3}^{2,0}(\bar{r})}{1-6gI_{2}(\bar{r})} \varepsilon_{B_{\mathrm{1g}}}. 
\end{align}
The denominator indicates the softening of $\phi$ when approaching the underlying (for $\alpha\varepsilon_{B_{\mathrm{1g}}}\! \!=\! 0$) nematic transition. This trend is also reflected in the susceptibility
\begin{align}
\tilde{\chi}_{\mathrm{nem}} = \frac{\partial\phi}{\partial\varepsilon_{B_{\mathrm{1g}}}}\bigg|_{\varepsilon_{B_{\mathrm{1g}}} \rightarrow 0} \!\!
=4\alpha\frac{J_{2}^{0,1}(\bar{r})-2J_{3}^{2,0}(\bar{r})}{1-6gI_{2}(\bar{r})}
\end{align}
which diverges at the transition temperature.

Let us note here that our analysis is performed at fixed \emph{strain}. The nematic transition hence corresponds to that observed in measurements of the elastoresistivity, with resistivity anisotropy $\Delta\rho$, as well as that obtained from the Raman response $R_{B_{\mathrm{1g}}} (\omega)$ in the $B_{\mathrm{1g}}$ channel \mycite{Gallais2013, Gallais2016, Karahasanovic2015}
\begin{align}
\tilde{\chi}_{\mathrm{nem}}\propto \frac{\partial\Delta\rho}{\partial\varepsilon_{B_{\mathrm{1g}}}}\bigg|_{\large \varepsilon_{B_{\mathrm{1g}}} \rightarrow 0} \!\! \propto\frac{2}{\pi}\int_{0}^{\infty}d\omega\frac{\mathrm{Im} [ R_{B_{\mathrm{1g}}} (\omega) ]}{\omega}.
\end{align}
This susceptibility $\tilde{\chi}_{\mathrm{nem}}$ differs from the 'true' thermodynamic nematic quantity
\begin{equation}
\chi_{\mathrm{nem}}= \frac{\partial\phi}{\partial  h_{\mathrm{nem}} }\bigg|_{T_{B_{\mathrm{1g}}} \rightarrow 0}=
\frac{6 I_{2}(\bar{r})}{1-6g_{\mathrm{ren}} I_{2}(\bar{r})},
\end{equation}
where $h_{\mathrm{nem}}$ is a field conjugate to the nematic order parameter and $g_{\mathrm{ren}\!} \!=\! g \!+\! \lambda_{\mathrm{n.e.}}/C^{(0)}_{66}$ is the nematic coupling constant renormalized by the coupling to elastic degrees of freedom \mycite{Fernandes2010}. Here $C^{(0)}_{66}$ is the bare (high-temperature) value of the elastic constant. The renormalized elastic constant is then $C_{66} \!=\! C_{66}^{(0)} / (1 + \lambda_{\mathrm{n.e.}}^{2} \chi_{\mathrm{nem}})$. The true susceptibility $\chi_{\mathrm{nem}}$ diverges at the nematic transition temperature for constant \emph{stress}. The fact that $g \!<\! g_{\mathrm{ren}}$ is the reason why the elastoresistivity or $\tilde{\chi}_{\mathrm{nem}}$ obtained from Raman measurements remain finite, and displays a temperature dependence $\tilde{\chi}_{\mathrm{nem}} \!\propto\! (T-\Theta_{\mathrm{C}})^{-1}$ with the Curie temperature $\Theta_{\mathrm{C}}$ below $T_{\mathrm{nem}}$ \mycite{Gallais2016, Karahasanovic2015}. The crucial distinction between these two susceptibilities was exploited in Ref.\ \mycite{Chu2012} to conclude that the origin of the nematic transition in the iron-based systems is electronic.

\subsubsection{$B_{\mathrm{2g}}$-strain}

The nematic symmetry is preserved for strain with the components $\varepsilon_{xy}\!=\!\varepsilon_{yx}\!=\!\varepsilon_{B_{\mathrm{2g}}}$. Here, the intra-plane inverse Green's functions \eqref{eq:inv-G} are simply augmented by the term $4 \varepsilon_{B_{\mathrm{2g}}} \delta (\vec{r} \!-\! \vec{r}')\partial_{x'}\partial_{y'}$. Now the self-consistency equations for $r$ and $\psi$ take the form
\begin{align}
\label{eq:self-consistency-eq-5}
r & =r_{0}+\int_{\vec{q}}\frac{6u(r+\vec{q}^{2}-4\varepsilon_{B_{\mathrm{2g}}} q_{x}q_{y})}{\mathcal{D}_{B_{\mathrm{2g}}} (\vec{q},\varepsilon_{B_{\mathrm{2g}}})} \\
\phi & =\int_{\vec{q}}\frac{6g\phi-2\alpha(q_{x}^{2}-q_{y}^{2})}{\mathcal{D}_{B_{\mathrm{2g}}} (\vec{q},\varepsilon_{B_{\mathrm{2g}}})},\label{eq:self-consistency-eq-6}
\end{align}
where 
\begin{align}
\mathcal{D}_{B_{\mathrm{2g}}} (\vec{q},\varepsilon_{B_{\mathrm{2g}}})
&= [r+\vec{q}^{2}-4\varepsilon_{B_{\mathrm{2g}}} q_{x}q_{y}]^{2} \\\nonumber
&\quad - [3g\phi-\alpha(q_{x}^{2}-q_{y}^{2})]^{2}
\end{align}
The systematic appearance of the strain $\varepsilon_{B_{\mathrm{2g}}}$ in combination with the odd function $q_{x}q_{y}$ in momentum space implies that all corrections to the unstrained case are (at least) quadratic in $\varepsilon_{B_{\mathrm{2g}}}$. An expansion similar to the earlier ones, with $r\approx\bar{r} + d\phi^{2} + d' \alpha^{2}+f_{B_{\mathrm{2g}}}\varepsilon_{B_{\mathrm{2g}}}^{2}$ provides
\begin{align}
f_{B_{\mathrm{2g}}} = f_{B_{\mathrm{1g}}}=4d'=\frac{24uJ_{3}^{2,0}(\bar{r})}{1+6uI_{2}(\bar{r})},
\end{align}
from Eq.\ \eqref{eq:self-consistency-eq-5}. The positive coefficient $f_{B_{\mathrm{2g}}}$ implies that the correlation length $\xi \!\sim\! (r - 3 g \phi)^{-1/2}$ is shortened by strain, pointing towards a generic \emph{suppression} of the nematic transition temperature. The equation of state \eqref{eq:self-consistency-eq-5} takes the form $0 = a_{B_{\mathrm{2g}}}\!(\varepsilon_{B_{\mathrm{2g}}})\phi + b\phi^{3}$, see Eq.\ \eqref{eq:effektiv-phi-hoch4}, with
\begin{align}\label{eq:GL-B2g}
a_{B_{\mathrm{2g}}}\!(\varepsilon_{B_{\mathrm{2g}}}) &= a + 6g\alpha^{2}[2 d' I_{3}(\bar{r})-K_{4}^{2,0}(\bar{r})] \\ \nonumber
 &\quad+ 12 g\varepsilon_{B_{\mathrm{2g}}}^{2}[f_{B_{\mathrm{2g}}} I_{3}(\bar{r}) - 6 K_{4}^{2,0}(\bar{r})].
\end{align}
Note that in the limit of quasi two-dimensional systems, the coefficient determining the shift of nematic transition is dominated by the logarithmic term in $J_{3}^{2,0}(\bar{r})$ (constituent of $d'$). Its magnitude is also decisive in determining the sign of the shift giving rise to a suppression of the nematic transition temperature.

\begin{figure}[tb]
\centering
\includegraphics[width = 0.45 \textwidth]{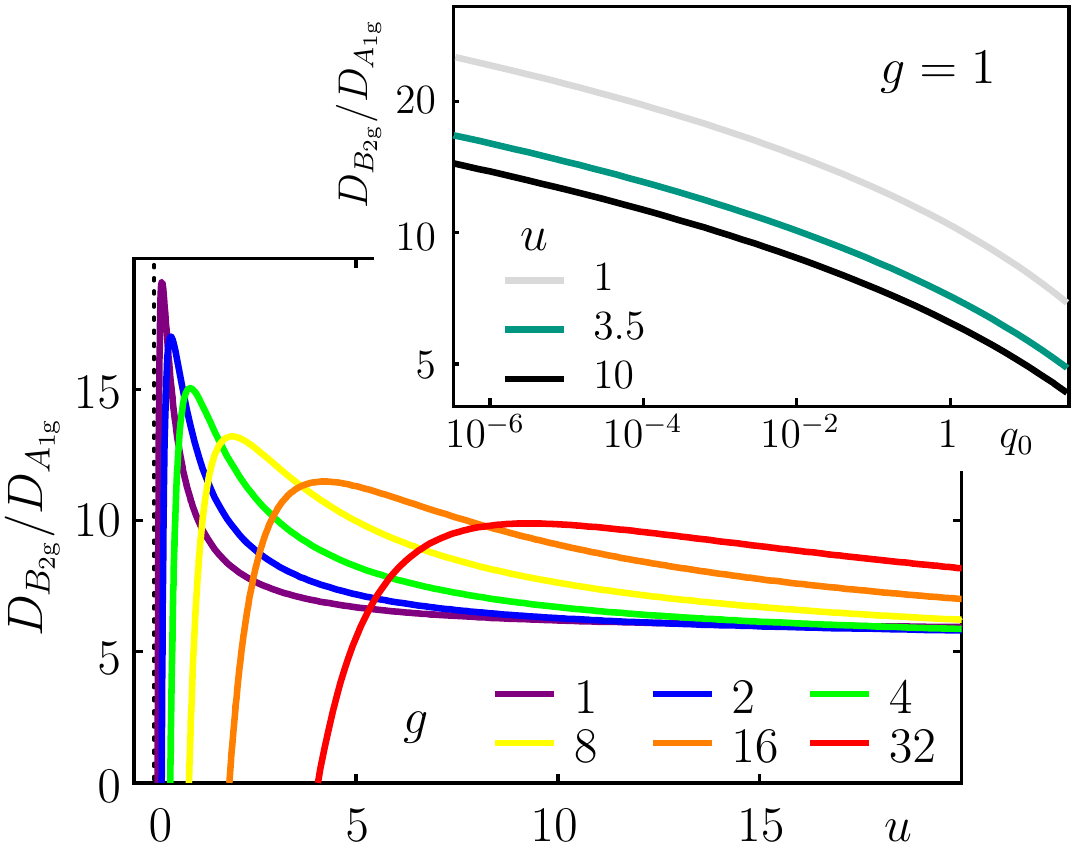}
\caption{Main: The quadratic departure of the nematic transition temperature upon applying strain depends on the strain's symmetry channel, and on material parameters. The ratio $D_{B_{\mathrm{2g}}}/D_{A_{\mathrm{1g}}}\!$---measuring the magnitude of this quadratic departure, see Eqs.\ \eqref{eq:B2g-expectation} and \eqref{eq:A1g-expectation}---is shown as a function of the phenomenological parameters $u$ for different values of $g$; while $q_{0} \!=\!1$ and $\Lambda \!=\! 10^{3}$ remain fixed. Inset: Here the same quantity is shown as a function of the anisotropy parameter $q_0$, for different valued of $u$; keeping $g = 1$ and $\Lambda = 10^{3}$ fixed. The values of the green curve are chosen close to the trictitical point found in the $\mathrm{Co}$-doped $\mathrm{Ba}\mathrm{Fe}_{2}\mathrm{As}_{2}$, see Ref.\ \mycite{Fernandes2012a}}
\label{fig:DoD-mat-params}
\end{figure}

\subsubsection{Discussion}

In the following discussion we focus on the (strongly) anisotropic limit $q_{0} \!\to\! 0$ relevant for many layered magnetic, or electronic systems. The Landau coefficients \eqref{eq:GL-A1g} and \eqref{eq:GL-B2g} can be expanded in the vicinity of the unstrained value $a$, for simplicity we assume $\alpha = 0$ here. This expansion requires using the expressions \eqref{eq:I-series-eval} and \eqref{eq:J-series-eval} listed in the Appendix \ref{app:integrals} as well as the solution $\bar{r} \approx 3 g q_{0} / 2\pi$ valid in the limit $q_{0} \ll g$. By associating the Landau coefficient with a nematic transition temperature via $a = (1 - T_{\mathrm{nem}}/T)$, we identify the phenomenological parameters
\begin{align}\label{eq:Db2g}
  D_{B_{\mathrm{2g}}} &= \Big(\frac{2u}{u+g}\Big)\ln\Big(\frac{\pi \Lambda^{2}}{3 g q_{0}}\Big) - \Big(\frac{5u+2g}{u+g}\Big),\\
  C_{A_{\mathrm{1g}}} &= -\Big(\frac{2u}{u+g}\Big), \;\text{and}\\
  D_{A_{\mathrm{1g}}} &= \Big(\frac{2u}{u+g}\Big)^{2},
\end{align} 
as defined in Eqs.\ \eqref{eq:B2g-expectation} and \eqref{eq:A1g-expectation}, and thereby quantify the effect of strain on the nematic phase transition.

For $A_{\mathrm{1g}}$ strain the quadratic correction is regular, while that of the $B_{\mathrm{2g}}$ channel is logarithmically divergent in the 2d limit. In light of the experimental finding \mycite{Ikeda2018}, that the response of the nematic phase boundary in $\mathrm{Co}$-doped $\mathrm{Ba}\mathrm{Fe}_{2}\mathrm{As}_{2}$ to the symmetric strain%
%
\footnote{The $B_{\mathrm{2g}}$ strain effects are to be mapped to the $B_{\mathrm{1g}}$ sector in the experimental work \mycite{Ikeda2018}, where the nematic axes point along the crystallographic diagonals (1,1) and (-1,1).}
$B_{\mathrm{2g}}$ as compared to that of the $A_{\mathrm{1g}}$ sector is very large, we compute the ratio $D_{B_{\mathrm{2g}}} / D_{A_{\mathrm{1g}}}$. Figure \ref{fig:DoD-mat-params} shows this ratio for fixed $q_{0}$ as a function of the material parameters $u$ and $g$, while the anisotropy-dependence for fixed $u$ and $g$ is shown as an inset. 
The figure covers a parameter range $-1 \lesssim \log(u/g) \lesssim 1$ and $q_{0} \ll 1$ in agreement with qualitative estimates for $\mathrm{Co}$-doped $\mathrm{Ba}\mathrm{Fe}_{2}\mathrm{As}_{2}$. Though a derivation of the phenomenological parameters from a full microscopic treatment exists, e.g.\ Ref.\ \mycite{Fernandes2012a}, the evaluation of an accurate numerical value remains difficult. As $g$ emerges from a perturbation theory \cite{Fernandes2012a}, the ratio $u/g \approx \{ 4 \mu m_{x} m_{y} / [\epsilon_{0} m (m_{x}-m_{y})] \}^{2}$ [with the chemical potential $\mu$, the offset energy $\epsilon_{0}$, and the hole ($m$) and electronic ($m_{x}$, $m_{y}$) band masses] is expected to be large, yet of order unity. Disorder effects \cite{Hoyer2016} manifestly increase $g$, hence decrease the ratio $u/g$. We estimate the anisotropy parameter $q_{0}$ for this pnictide compound from $q_{0}^{2} \sim J_{ab} / J_{c} \approx 30$ as reported in Ref.\ \mycite{Pengcheng2015}.

The representations in Fig.\ \ref{fig:DoD-mat-params} support that the $B_{\mathrm{2g}}$ strongly affects the nematic order; possibly to the point of reaching beyond linear effects $\propto C_{A_{\mathrm{1g}}} \varepsilon_{A_{\mathrm{1g}}}$ from the $A_{\mathrm{1g}}$ sector. Although derived in the strongly anisotropic limit, $q_{0} \!\to\! 0$, Eq.\ \eqref{eq:Db2g} points towards a sign change of $D_{B_{\mathrm{2g}}}$ when moving away from the two-dimensional limit; a possibility that depends on the specific parameters in the problem.

\section{Conclusion}
In conclusion, we have investigated the response of a nematic order to strain belonging to different symmetry classes ($\varepsilon_{A_{\mathrm{1g}}}$, $\varepsilon_{B_{\mathrm{1g}}}$, $\varepsilon_{B_{\mathrm{2g}}}$). Hereby special attention was given to the symmetry channel $B_{\mathrm{2g}}$, for which recent experimental work \mycite{Ikeda2018} has found a surprisingly strong suppression of the nematic transition temperature $T_{\mathrm{nem}}$. Our analysis of long- and short-length fluctuations in a $J_{1}$-$J_{2}^{\pm}$ spin model---the latter implements a $B_{\mathrm{2g}}$-strain on a lattice---provides us with clear indications that the nematic transition temperature \emph{decreases} with strain. This is due to the strain-induced shortening of the magnetic correlation length $\xi$.

These findings, combined with more general symmetry considerations lead to several observations. The degeneracy-lifting $B_{\mathrm{1g}}$ strain is expected to act as an effective source field $h_{\mathrm{eff}}$, thus replacing the nematic transition by a smooth cross-over. In contrast, the symmetry-conserving strains of the $A_{\mathrm{1g}}$ and $B_{\textrm{2g}}$ type preserve the transition and merely yield a shift $\Delta T_{\mathrm{nem}} = T_{\mathrm{nem}}(\varepsilon) - T_{\mathrm{nem}}^{0}$ in the transition temperature. In the $A_{\mathrm{1g}}$ channel, we find $\Delta T_{\mathrm{nem}} = C_{A_{\mathrm{1g}}} \varepsilon_{A_{\mathrm{1g}}} - D_{A_{\mathrm{1g}}} \varepsilon_{A_{\mathrm{1g}}}^{2}$, with a dominant linear contribution. In the $B_{\mathrm{2g}}$ sector, the deviation is quadratic, i.e.\  $\Delta T_{\mathrm{nem}} = - D_{B_{\mathrm{2g}}} \varepsilon_{B_{\mathrm{2g}}}^{2}$.

Approaching the problem from a field-theoretical approach we provide a quantitative tool to evaluate the parameters $h_{\mathrm{eff}}$, $C_{A_{\mathrm{1g}}}$, $D_{A_{\mathrm{1g}}}$ and $D_{B_{\mathrm{2g}}}$ from the underlying Lagrangian formalism. In this approach, the strong response of nematic order to symmetric $B_{\mathrm{2g}}$, i.e.\ $D_{B_{\mathrm{2g}}} \gg D_{A_{\mathrm{1g}}}$, is found to be a generic feature of strongly anisotropic, quasi-2d systems, see Fig.\ \ref{fig:Tnem-evolution}; as represented by $\mathrm{Co}$-doped $\mathrm{Ba}\mathrm{Fe}_{2}\mathrm{As}_{2}$.

A related issue is that other emergent vestigial phases are known to appear in the spin-based scenario of iron-based materials \mycite{Fernandes2016, Fernandes2018-arXiv}. Furthermore, charge density wave order is expected to feature similar responses in a situation of $C_{4}$-symmetric collinear double-$\vec{Q}$ spin order; as observed in a number of systems \mycite{Kim2010, Hassinger2012,  Avci2014, Wang2016, Boehmer2015, Allred2015, Hassinger2016, Allred2016}. Finally, chiral order is related to $C_{4}$-symmetric non-collinear double-$\vec{Q}$ spin order that forms spin vortex crystals observed in Ref.\ \mycite{Meier2018}. This embeds our work in a broader quest for understanding how strain affects vestigial phases of matter.

\begin{acknowledgments}
We are grateful to I. R. Fisher, M. Ikeda, J. C. Palmstrom, and P.
Walmsley, for stimulating discussions. J.\ S.\  was funded by the Gordon
and Betty Moore Foundation\textquoteright s EPiQS Initiative through
Grant GBMF4302 while visiting the Geballe Laboratory for Advanced
Materials at Stanford University.
\end{acknowledgments}

\vfill
\pagebreak

\bibliography{strain-tuning}

\onecolumngrid
\pagebreak
\hfill {\large \textbf{Appendix}}\hfill\phantom{emphty}\\
\twocolumngrid

\appendix

\section{Momentum Integrals}
\label{app:integrals}
The derivation of the results in the main text involve to compute several integrals of the form $\int_{{\vec{q}}}f(\bar{r},{\vec{q}})$, which we define in the following as 
\begin{align}
I_{n}(r) & \equiv\int_{{\vec{q}}}\frac{1}{(r+{\vec{q}}^{2})^{n}}\label{eq:I-series}\\
J_{n}^{\ell,m}(r) & \equiv\int_{{\vec{q}}}\frac{(q_{x}^{2}-q_{y}^{2})^{\ell}(q_{x}^{2}+q_{y}^{2})^{m}}{(r+{\vec{q}}^{2})^{n}}\label{eq:J-series}\\
K_{n}^{\ell,m}(r) & \equiv\int_{{\vec{q}}}\frac{(q_{x}^{2}-q_{y}^{2})^{\ell}(q_{x}q_{y})^{m}}{(r+{\vec{q}}^{2})^{n}}.\label{eq:K-series}
\end{align}
Angular integration in the $xy$ plane provides the relations
\begin{align}
J_{n}^{\ell,m}(r) & = \frac{\mathrm{B}(\ell)\ell!}{(\ell!!)^{2}}J_{n}^{0,\ell+m}(r),\\
K_{n}^{\ell,m}(r) & = \frac{\mathrm{B}(\ell)\mathrm{B}(m)\ell!\,m!}{2^{m}(\ell!!)(m!!)((\ell+m)!!)}J_{n}^{0,\ell+m}(r),
\end{align}
with $z!\equiv\Gamma(z+1)$ and $(2z)!!=2^{z}z!$ defined through
the Gamma function $\Gamma(\zeta)\equiv\int_{0}^{\infty}dt\,t^{\zeta-1}e^{-t}$
and $\mathrm{B}(z)=[1+(-1)^{z}]/2$ the Boolean parity function. Furthermore,
we have $J_{n}^{0,0}(r)=K_{n}^{0,0}(r)=I_{n}(r)$. The remaining task
is to deterine the integrals $I_{n} (r)$.

\subsection*{Integrals in the Anisotropic 3d limit}

\label{app:anis-3d} Next we evaluate the integrals $I_{n} (r)$
in the anisotropic 3d case. We substitute $q_{x}\!\to\!q\cos(\varphi)$,
$q_{y}\!\to\!q\sin(\varphi)$, and $q_{z}^{2}\!\to\!2q_{0}^{2}[1-\cos(q_{z}/q_{0})]$
with $q_{0}$ a measure of the uniaxial anisotropy; now the momentum-space
integration along $z$ is restricted to $|q_{z}|<\pi/q_{0}$. With
an ultraviolet cut-off $\Lambda$ for the in-plane momentum $q=(q_{x}^{2}+q_{y}^{2})^{1/2}$,
$q<\Lambda$, the momentum-integral are mapped to 
\begin{align}
\int_{\vec{q}}f(r,\vec{q})\to\int\limits_{-\pi q_{0}}^{\pi q_{0}}\frac{dq_{z}}{2\pi}\int\limits_{0}^{\Lambda}\frac{qdq}{2\pi}\int\limits_{0}^{2\pi}\frac{d\varphi}{2\pi}f(r,\vec{q}).
\end{align}

Within this mapping, the integrals defined in Eq.\ \eqref{eq:I-series}
evaluate to 
\begin{align}
I_{1}(r) & = \frac{q_{0}}{4\pi}\Big[\ln\Big(\frac{\Lambda^{2}}{q_{0}^{2}}\Big)-\ln\Big(\frac{r+2q_{0}^{2}+\sqrt{r(r+4q_{0}^{2})}}{2q_{0}^{2}}\Big)\Big],\nonumber \\
I_{2}(r) & = \frac{1}{4\pi}\frac{q_{0}}{[r(r+4q_{0}^{2})]^{1/2}},\nonumber \\
I_{3}(r) & = \frac{1}{8\pi}\frac{q_{0}(r+2q_{0}^{2})}{[r(r+4q_{0}^{2})]^{3/2}},\nonumber \\
I_{4}(r) & = \frac{1}{12\pi}\frac{q_{0}(r^{2}+4rq_{0}^{2}+6q_{0}^{4})}{[r(r+4q_{0}^{2})]^{5/2}},\nonumber \\
\label{eq:I-series-eval}
I_{n} (r) & = {}_{2}F_{1} \Big[1/2,n-1,1,\frac{4q_{0}^{2}}{4q_{0}^{2}+r}\Big]q_{0},
\end{align}
with $_{2}F_{1}$ the hypergeometric function. The limit $\Lambda\!\to\!\infty$
is taken for all convergent integrals, while only the dominant terms
in $\Lambda$ are considered otherwise.

Below, we evaluate some integrals of the form \eqref{eq:J-series}
and express them in terms of the series of $I_{n}$ functions. 
\begin{align}
J_{3}^{2,0}(r) & = \frac{J_{3}^{0,2}(r)}{2}=\frac{I_{1}(r)}{2}-\frac{3q_{0}}{16\pi},\nonumber \\
J_{4}^{2,0}(r) & = \frac{J_{4}^{0,2}(r)}{2}=\frac{I_{2}(r)}{6},\nonumber \\
J_{2}^{0,1}(r) & = I_{1}(r)-\frac{q_{0}}{4\pi},\nonumber \\
J_{3}^{0,1}(r) & = \frac{I_{2}(r)}{2},\nonumber \\
J_{4}^{0,1}(r) & = \frac{I_{3}(r)}{3}.
\end{align}
More generically we find 
\begin{align}
J_{n}^{0,m}(r) & = \frac{I_{n-m}(r)}{C(n-1,m)},\qquad\text{for }n-m\geq2\nonumber \\
J_{n}^{0,n-1}(r) & = I_{1}(r)-\frac{c_{n}q_{0}}{4\pi}.\label{eq:J-series-eval}
\end{align}
with $C(n-1,m)=(n-1)!/[(n-m-1)!m!]$ the binomial coefficients, $c_{n}=\sum_{k=1}^{n-1}(1/k)=\Gamma'(n)/\Gamma(n)+\gamma_{{\scriptscriptstyle \mathrm{E}}}$
positive constants related to the Polygamma function, and $\gamma_{{\scriptscriptstyle \mathrm{E}}}\approx0.577$
Euler's constant.

\onecolumngrid
\vfill

\end{document}